\documentclass[12pt]{article}
\usepackage{amsmath}
\usepackage{times}
\usepackage{graphicx}
\usepackage{color}
\usepackage{multirow}
\usepackage{rotating}
\usepackage{bbm}
\usepackage{latexsym}

\usepackage{amsmath}
\usepackage{amssymb,bm}
\usepackage{graphicx}
\usepackage{color}
\usepackage[dvipsnames]{xcolor}
\makeatletter
\newif\if@restonecol
\makeatother

\usepackage{algorithm}
\usepackage{algpseudocode}
\DeclareMathAlphabet\mathpzc{OT1}{pzc}{m}{it}
\let\mathcal=\mathpzc

\def\H{{\mathbb H}}

\let\<=\langle
\let\>=\rangle

\def\~#1{{\text{\sf#1}}}

\let\-=\mathbf
\let\^=\hat

\def\circ{\ifmmode\mathchar"220E\else$\mathchar"220E$\fi}
\def\@#1{{\cal #1}}

\def\COV{\mathrm{COV}}
\newcommand{\captionfonts}{\normalsize}

\makeatletter  
\long\def\@makecaption#1#2{%
  \vskip\abovecaptionskip
  \sbox\@tempboxa{{\captionfonts #1: #2}}%
  \ifdim \wd\@tempboxa >\hsize
    {\captionfonts #1: #2\par}
  \else
    \hbox to\hsize{\hfil\box\@tempboxa\hfil}%
  \fi
  \vskip\belowcaptionskip}
\makeatother   

\begin{document}
\hspace{13.9cm}1

\ \vspace{20mm}\\

{\LARGE Adaptive Gaussian process approximation for Bayesian inference with expensive likelihood functions}

\ \\
{\bf \large Hongqiao Wang$^{\displaystyle 1}$ and Jinglai Li$^{\displaystyle 2}$}\\
{$^{\displaystyle 1}$Institute of Natural Sciences and School of Mathematical Sciences, Shanghai Jiao Tong University,
Shanghai 200240, China.}\\
{$^{\displaystyle 2}$Institute of Natural Sciences, School of Mathematical Sciences, and MOE Key Laboratory of Scientific and Engineering Computing, Shanghai Jiao Tong University, Shanghai 200240, China. (Corresponding author).}\\
%


\thispagestyle{empty}
\markboth{}{NC instructions}
\ \vspace{-0mm}\\
%
\begin{center} {\bf Abstract} \end{center}
We consider Bayesian inference problems with computationally intensive likelihood functions.
We propose a Gaussian process (GP) based method to approximate the joint distribution of the unknown parameters and the data, built upon a recent work~\cite{kandasamy2015bayesian}. 
In particular, we write the joint density approximately as a product of an approximate posterior density and an exponentiated GP surrogate. 
We then provide an adaptive algorithm to construct such an approximation, where an active learning method is used to choose the design points. 
With numerical examples, we illustrate that the proposed method has competitive performance against existing approaches for Bayesian computation. 

\section{Introduction}\label{s:intro}

The Bayesian inference is a popular method to estimate unknown parameters from data,
and a major advantage of the method is its ability to quantify uncertainty in the inference results~\cite{gelman2014bayesian}. 
In this work we consider Bayesian inference problems where the likelihood functions are highly expensive to evaluate.
A typical example of this type of problems is the Bayesian inverse problems~\cite{tarantola2005inverse}, where the parameters of interest can not be observed directly and need to be
estimated from indirect data.
Such problems arise from many real-world applications, ranging from carbon capture~\cite{konomi2016bayesian} to chemical kinetics~\cite{golightly2011bayesian}.
In  Bayesian inverse problems, the mappings from the parameter of interest to the observable quantities, often known as the forward models, are often computationally intensive, e.g., involving simulating large scale computer models.

Due to the high computational cost, common numerical implementations of Bayesian inferences, such as the Markov chain Monte Carlo (MCMC)~\cite{andrieu2003introduction}  methods can be prohibitively expensive. 
A simple idea to accelerate the computation of the posterior is to construct a computationally inexpensive surrogate or an approximation of the posterior distribution
with a limited number of likelihood function evaluations. 
To this end, a particular convenient choice for surrogate function is the Gaussian Process~(GP) model~\cite{williams2006gaussian}. 
The idea of using the GP model to approximate the posterior or the likelihood function dates back to the the so-called Bayesian quadrature (or Bayesian Monte Carlo) approaches~\cite{o1991bayes,rasmussen2003bayesian,rasmussen2003gaussian,kennedy1998bayesian}, which were designed to perform numerical integrations in a Bayesian fashion (for example, to compute the evidence in Bayesian inference problems~\cite{osborne2012active}). 
Unlike the Bayesian quadrature methods, the goal of this work is to construct an approximation of the posterior distribution. 
To this end,  a recent work~\cite{kandasamy2015bayesian} approximates the joint distribution of the unknown parameter and the data (which can also be viewed as the un-normalized posterior distribution) with an
exponentiated GP model, where the design points, i.e., the points where the likelihood function is evaluated, are chosen with an active learning strategy.
In particular, they determine the the design points  by sequentially maximizing the variance in the posterior approximation.  
{Other ideas of using the GP approximation to accelerate the Bayesian computation can be found in \cite{conrad2016accelerating, bilionis2013solution}, and so on. }
The method presented in this work also intends to approximate the un-normalized posterior distribution. 
The main contribution of the work is the following. 
{  We write the unnormalized posterior distribution as a product of an approximate posterior density and an exponentiated GP surrogate. 
The intuition behind this formulation is that,  the GP model can be more effectively constructed if we factor out a good approximation of the posterior (see Section~\ref{sec:alg} for a detailed explanation). 
As we may not know a good approximate posterior density in advance, we develop an algorithm to adaptively construct the product-form approximation of the un-normalized posterior distribution. }
{Another difference between our method can that in \cite{kandasamy2015bayesian} is the learning strategy for selecting the design points. 
Namely, we use the entropy rather than the variance as the selection criterion, which can better represent the uncertainty in the approximation. }
Numerical examples illustrate that the proposed method can substantially improve the performance of the GP approximation. 

{ We note that, other surrogate models, notably the generalized polynomial chaos~(gPC) expansion~\cite{li2014adaptive,marzouk2009stochastic,marzouk2009dimensionality,marzouk2007stochastic,nagel2016spectral}, have also been used to accelerate the Bayesian computation. 
Detailed comparison of  the two type of the surrogates is not discussed in this work and those who are interested in this matter may consult  \cite{o2013polynomial}.}

The rest of the paper is organized as the following. In Section~\ref{sec:agp} we present the adaptive GP algorithm to construct the posterior approximation and the active leaning method to determine the design points. 
In section~\ref{sec:examples}, we give two examples to illustrate the performance of the proposed method. 
Finally section~\ref{sec:conclusions} provides some concluding remarks.

\section{The adaptive GP method} \label{sec:agp}
\subsection{Problem Setup}
A Bayesian inference problem aims to estimate an unknown parameter $\-x$ from data $\-d$, and specifically it computes the posterior distribution
of $\-x$ using the Bayes' formula:  
\begin{equation}
\pi(\-x|\-d) \propto\pi(\-x,\-d)={l(\-d|\-x)\pi(\-x)},
\end{equation}
where $l(\-d|\-x)$ is the likelihood function and $\pi(\-x)$ is the prior distribution of $\-x$.
When the Bayesian method is applied to inverse problems, the data and the forward model enter the formulation through the likelihood function.
Namely, suppose that there is a function (termed as the forward function or the forward model) that maps the parameter of interest $\-x$ to the observable quantity $\-y$: \[\-y = \-G(\-x)+\-z,\]
where $\-z$ is the observation error. Now we further assume that the distribution density of the observation noise $\-z$, $p_z(\-z)$, is available, and it follows directly that the likelihood function is given by \[l(\-d|\-x) = p_z(\-d-\-G(\-x)).\] 
In what follows we shall omit the argument $\-d$ in the likelihood function and denote it as $l(\-x)$ for simplicity. 
It is easy to see that each evaluation of the likelihood function $l(\-x)$ requires to evaluate the forward function $\-G(\-x)$. 
In practice, the forward function $\-G(\-x)$ often represents a large-scale computer model, and
 thus the evaluation of $l(\-x)$ can be highly computational demanding.
 Due to the high computational cost,  the brute-force Monte Carlo simulation can not be used for such problems, 
 and we resort to an alternative method to compute the posterior distributions, using the GP surrogate model. A brief description of the GP method is provided in next section.  
 
 \subsection{The GP model}
\label{sec:gp}
Given a real-valued function $g(\-x)$,  the GP or the Kriging method constructs a surrogate model of $g(\-x)$ in a nonparameteric Bayesian regression framework~\cite{williams2006gaussian,oakley2002bayesian,OHagan1978}.
Specifically the target function $g(\-x)$ is cast as a Gaussian random process 
whose mean is $\mu(\-x)$ and covariance is specified by
a kernel function $k(\-x,\-x')$, namely,
\[ \COV[g(\-x),g(\-x')] = k(\-x,\-x'). \]
The kernel $k(\-x,\-x')$ is positive semidefinite and bounded. Now let us assume that $m$ evaluations of the function $g(\-x)$ are performed
 at parameter values $\-X^* := \left[\-x^*_1, \ldots \-x^*_m\right]$, yielding  function evaluations $\-y^* := \left[ {y}^*_1, \ldots {y}^*_m\right]$,
where \[{y}^*_i = g(\-x_i^*)\quad \mathrm{for} \quad i=1,\ldots,m.\]
Suppose that we want to predict the function values at points $\-D := \left[\-x_1, \ldots \-x_{m'}\right]$, i.e., $\-y=[y_1,\ldots y_{m'}]$ where
$ {y_i=g(\-x_i)}$. The sets $\-X^*$ and $\-D$ are often known as the training and the test points respectively. 
The joint prior distribution of $(\-y^*,\,\-y)$ is, 
\begin{equation}  
\left[ \begin{array}{c}
         \-y^* \\
        \-y \end{array} \right] \sim \@N\left(\begin{array}{c}
         \mu(\-X^*) \\
        \mu(\-D) \end{array},
				\left[
				\begin{array}{ll}
         K(\-X^*,\-X^*) &K(\-X^*,\-D) \\
        K(\-D,\-X^*) &K(\-D,\-D) \end{array}\right]
				\right) , \label{e:jointdis}
\end{equation}
where we use the notation $K(\-A,\-B)$ to denote the matrix of the covariance evaluated at all pairs of points in set $\-A$ and in set $\-B$.
The posterior distribution of $\-y$ is also Gaussian: 
\begin{subequations}
\label{e:gp}
\begin{equation}
  \-y ~|~\-D,\-X^*, \-y^* \sim\mathcal{N}(\-u, \Sigma), \label{e:post}
\end{equation}
where the posterior mean is 
\begin{equation}
\-u=\mu(\-D)+K(\-D,\-X^*)K(\-X^*,\-X^*)^{-1}(\-y-\mu(\-D)),
\end{equation}
and the posterior covariance matrix is 
\begin{equation}
\label{eq:postcovsimple}
\Sigma = K(\-D,\-D)-K(\-D,\-X^*)K(\-X^*,\-X^*)^{-1}K(\-X^*,\-D). 
\end{equation}
\end{subequations}
Here we only provide a brief introduction to the GP method tailored for our own purposes, and readers who are interested in further details may consult the aforementioned references.  


\subsection{ The adaptive GP algorithm}\label{sec:alg}
Now we discuss how to use the GP method to compute the posterior distribution in our problem. 
A straightforward idea is to construct the surrogate model directly for the log-likelihood function $\log l(\-x)$, and such a method 
has been used in the aforementioned works~\cite{osborne2012active,kandasamy2015bayesian}. 
A difficulty in this approach is that the target function $\log l(\-x)$ can be highly nonlinear and fast varying, and thus are not well described by a GP model. 
We here present an adaptive scheme to alleviate the difficulty.

We first write the unnormalized posterior, i.e., the joint distribution $\pi(\-x,\-d)$, as
\[f(\-x)=l(\-x)\pi(\-x)=\exp({g}(\-x) )p(\-x) ,\]
where $p(\-x)$ is a probability distribution that we are free to choose and 
 \begin{equation}
 g(\-x)=\log(f(\-x)/p(\-x)). \label{e:gx}
 \end{equation} 
We work on the log posterior distribution since the log smoothes out a function and is more conducive for the GP modeling. 
 Also, by doing this we ensure the non-negativity of the obtained approximate posterior. 
We then sample the function $g(\-x)$ at certain locations and construct the GP surrogate of $g(\-x)$.  
It should be noted that, the distribution $p(\-x)$ plays an important role in the surrogate construction as a good choice of $p(\-x)$
can significantly improve the accuracy of the GP surrogate models. 
In particular, if we take $p(\-x)$ to be exactly the posterior $\pi(\-x|\-d)$, it follows immediately that $g(\-x)$ in Eq~\eqref{e:gx} is a constant. 
This then gives us the intuition that, if $p(\-x)$ is a good approximation to the posterior distribution $\pi(\-x|\-d)$, $g(\-x)$ is a mildly varying function which
is easy to approximate. 
In other word, we can improve the performance of the GP surrogate by factoring out a good approximation of the posterior. 
Certainly, this can not be done in one step, as the posterior is not known in advance.
We present here an adaptive framework to construct a sequence of pairs $\{p_i(\-x), \exp(\hat{g}_i(\-x))\}$, the product of which evolves to 
a good approximation of the unnormalized posterior $f(\-x)$.  
Roughly speaking the algorithm performs the following iterations:  in the $n$-th cycle, given the current guess of the posterior distribution $p_n(\-x)$, we construct 
a GP surrogate $\hat{g}_n(\-x)$ of $g_n(\-x)$ which is given by 
\begin{equation*}
 g_n(\-x)=\log(f(\-x)/p_n(\-x)), 
 \end{equation*} 
and we then compute a new (and possibly better) posterior approximation $p_{n+1}(\-x)$ using 
\[p_{n+1}(\-x) \propto \exp(\hat{g}_n(\-x))p_{n}(\-x).\]
{Finally we want to specify stopping criteria for the iteration, and the iteration terminates if either of the following two conditions are satisfied.  
The first is that the maximum number of iterations is reached. Our second stopping condition is based on the Kullback-Leiber (KL) divergence between $p_{n-1}$ and $p_n$, which 
reads,
\begin{equation}
D_{KL}(p_{n-1}, p_n) = \int \log\frac{p_{n-1}(\-x)}{p_n(\-x)} p_{n-1}(\-x) d\-x.
\end{equation}
Specifically the second stoping condition is that $D_{KL}(p_{n-1}, p_n) $ is smaller than a prescribed value $D_{\max}$ in $K$ consecutive iterations. 
That is, if the computed posterior approximation does not change much in a certain number of consecutive iterations, the algorithm terminates.}
The complete scheme is described in Algorithm~\ref{alg:ak}.
\begin{algorithm}
    \caption{The adaptive GP algorithm}
    \label{alg:ak}
    \begin{algorithmic}[1]
    \Require{$m_0$, $n_{\max}$, $M$, $D_{\max}$, $k_{\max}$}
	
       \State let $\hat{p}_0(\-x) = \pi(\-x)$;  let $n=0$; let $k=0$;
         \State choose $m_0$ initial design points: $\{x_1,...,x_{m_0}\}$, and compute $y_i=f(\-x_i)$ for $i=1...m_0$;
         \State let $S_0 = \{(x_1,y_1),\,...,\,(x_{m_0},y_{m_0})\}$;

  \For{n=0} {$n_{\max}$}
				   \State let $g_n(\-x) = \log (f(\-x)/\hat{p}_n(\-x))$;
	\State construct a GP surrogate model $\hat{g}_n(\-x)$ for the function $g_n(\-x)$ with data set $S_n$; \label{l:gp}
	\State draw a set of $M$ samples from the approximate posterior \[p_{n+1}(\-x)\propto\exp(\hat{g}_n(\-x))\hat{p}_n(\-x)\]
	 with MCMC, denoted as $A_n$; \label{l:mcmc}
	\State obtain an estimated PDF from samples $A_n$, denoted as $\hat{p}_{n+1}$; \label{l:de}
	\State compute $D_{KL}(\hat{p}_{n-1}, \hat{p}_{n})$; \label{l:kld}
	\If{$D_{KL}(\hat{p}_{n-1}, \hat{p}_{n})< D_{\max}$}  {$k = k+1$;}
	 \Else 
	 { $k=0$;}
	\EndIf
	 \If{$k = K$} {break the FOR loop;}
	\Else 
	\State select $m$ design points: $\{x_1,...,x_{m}\}$, evaluate $f(\-x_i)$ for $i=1...m$, and  
	let $S_{n+1} = S_n \cup \{(x_1,\,y_1),...,(x_{m},y_{m})\}$; \label{l:design}
\EndIf	
	\EndFor
       
   \end{algorithmic}
\end{algorithm}

Some remarks on the implementation of Algorithm~\ref{alg:ak} are listed in order:
\begin{itemize}
\item In Line~\ref{l:gp}, we construct the GP model for $g_n(\-x)$ using the procedure described in Section~\ref{sec:gp}.
{The hyperparameters of the GP model are determined by maximizing the marginal likelihood function~\cite{williams2006gaussian}. }
\item In Line~\ref{l:mcmc}, we  resort to the MCMC method to draw a rather large number of samples from the approximate posterior distribution; this procedure, however, does not require to evaluate the true likelihood function and is not computationally expensive. 
\item In Line~\ref{l:de}, we need to compute  the density function of a distribution $p_{n+1}$ from the samples $X_n$, and here we use the Gaussian mixture method~\cite{mclachlan2004finite} to 
estimate the density.
Certainly there will be estimation errors in this procedure and so we denote the estimated density as $\hat{p}_{n+1}$ to distinguish it from the true density $p_{n+1}$. 
\item { In Line~\ref{l:kld}, we find that  it is rather costly to compute the KLD between $p_{n-1}$ and $p_{n}$.
We instead use the KLD between $\hat{p}_{n-1}$ and $\hat{p}_n$, which is much easier to compute as the distributions are available as Gaussian mixtures.}
\item In Line~\ref{l:design}, we need to determine the design points, i.e., the locations where we evaluate the true function.  The choice of design points is
critical to the performance of the proposed adaptive GP algorithm, and we use an active learning method to determine the points, which is presented in Section~\ref{sec:entropy}. 
\end{itemize}
\subsection{Active learning for the design points}
\label{sec:entropy}
In the GP literature, 
the determination of the design points is often cast as an experimental design problem, i.e., to find the experimental parameters that can provide 
us the most information. 
The problem has received considerable attention and a number of methods and criteria have been proposed to select the points,
such as, the Mutual Information criterion~\cite{krause2008near},
the Integrated Mean Square Error (IMSE)~\cite{sacks1989design}, the Integrated Posterior
Variance (IVAR)~\cite{gorodetsky2016mercer}, and the active learning MacKay (ALM) criterion~\cite{mackay1992information},  just to name a few. 
Here we choose to use an active learning  strategy, that adds one design point a time, primarily for that it is easy to implement. 

A common active learning strategy is to choose the point that has the largest uncertainty, and to this end we need a function that can measure or quantify the uncertainty in the approximation reconstructed. 
In the usual GP problems, the variance of the GP model $\hat{g}(\-x)$ is a natural choice for such a measure of uncertainty (which yields 
the ALM method), because the distribution of $\hat{g}(\-x)$ is Gaussian.
In our problems, however,  the function of interest is the posterior approximation $\hat{f}(\-x)=\exp(\hat{g}(\-x))p(\-x)$ rather than
the GP model $\hat{g}(\-x)$ itself, and thus we should measure the uncertainty in $\hat{f}(\-x)$. 
In \cite{kandasamy2015bayesian}, the variance of the posterior approximation $\hat{f}$ is used as the measure function. 
However, since the distribution of $\hat{f}(\-x)$ is not Gaussian, the variance may not provide a good estimate of the uncertainty. 
On the other hand, the entropy is a commonly used  measure to quantify the uncertainty in a random variable~\cite{renyi1961measures,shannon2001mathematical}, and
here we use it as our design criterion. 

Specifically, suppose that,  at point $\-x$, the distribution of $\hat{f}(\-x)$ is $\pi_f(\hat{f})$, and the entropy of $\hat{f}(\-x)$ is defined as
\begin{equation}
\H(\hat{f}(\-x)) = -\int \log(\pi_f(\hat{f})) \pi_f(\hat{f}) d\hat{f}.\label{e:entropy}
\end{equation}
Thus we choose a new design point by 
\[\max_{\-x\in \Omega} \H(\hat{f}(\-x)),\] 
where $\Omega$ is a bounded subspace of the state space of $\-x$.  
In the present problem, the distribution of $\hat{g}(\-x)$ is Gaussian and let us assume its mean and variance are $\mu$ and $\sigma^2$ respectively. 
It follows that the distribution of $\hat{f}(\-x)$ is log-normal and the entropy of it can be computed analytically:
\begin{equation}
\H(\hat{f}) = \mu+\frac1\ln(2\pi e\sigma^2). \label{e:hlog}
\end{equation}
We want to emphasize that the entropy based active learning method is different from the usual maximum entropy method for experimental design, e.g., \cite{sebastiani2000maximum}.
The purpose of the maximum entropy method in \cite{sebastiani2000maximum}, is the find the design points that maximize the information gain of an inference problem,
 while in our problem, we use the entropy as a measure of uncertainty. 

Now suppose that we have a set of existing data points, and we want to choose $m$ new design points. 
We use the following scheme to sequentially choose the new points:
\begin{enumerate}
\item Construct a GP model $\hat{g}(\-x)$ for $g(\-x)$ using data set $S$; \label{st:gp}
\item Compute $\-x^* = \arg\max_{\-x\in\Omega} \H(\hat{f}(\-x))$; \label{st:maxent}
\item Evaluate $y^* = g(\-x^*)$ and let $S=S\cup \{(\-x^*,y^*)\}$; 
\end{enumerate}
Note that the key in the adaptive scheme is Step~\ref{st:maxent}, where we seek the point $\-x$
that maximizes the entropy $\H(\hat{f}(\-x))$ in $\Omega$. 
This is a quite challenging problem from an optimization perspective, because 
{the gradient of the objective function can not be easily obtained and }
the problem may have multiple local maxima. 
However, in the numerical tests, we have found that, our algorithm does not strictly require the optimality of the solution and it performs well as long as 
a good design point can be found in each step.  Thus here we use a stochastic search method, the simulated annealing algorithm~\cite{kirkpatrick1983optimization},  to find the design point. 
We have also tested other meta-heuristic optimization algorithms,  and the performances do not vary significantly. 



\section{Numerical examples} \label{sec:examples}

\subsection{The Rosenbrock function}
\begin{figure}
\centerline{\includegraphics[width=.5\textwidth]{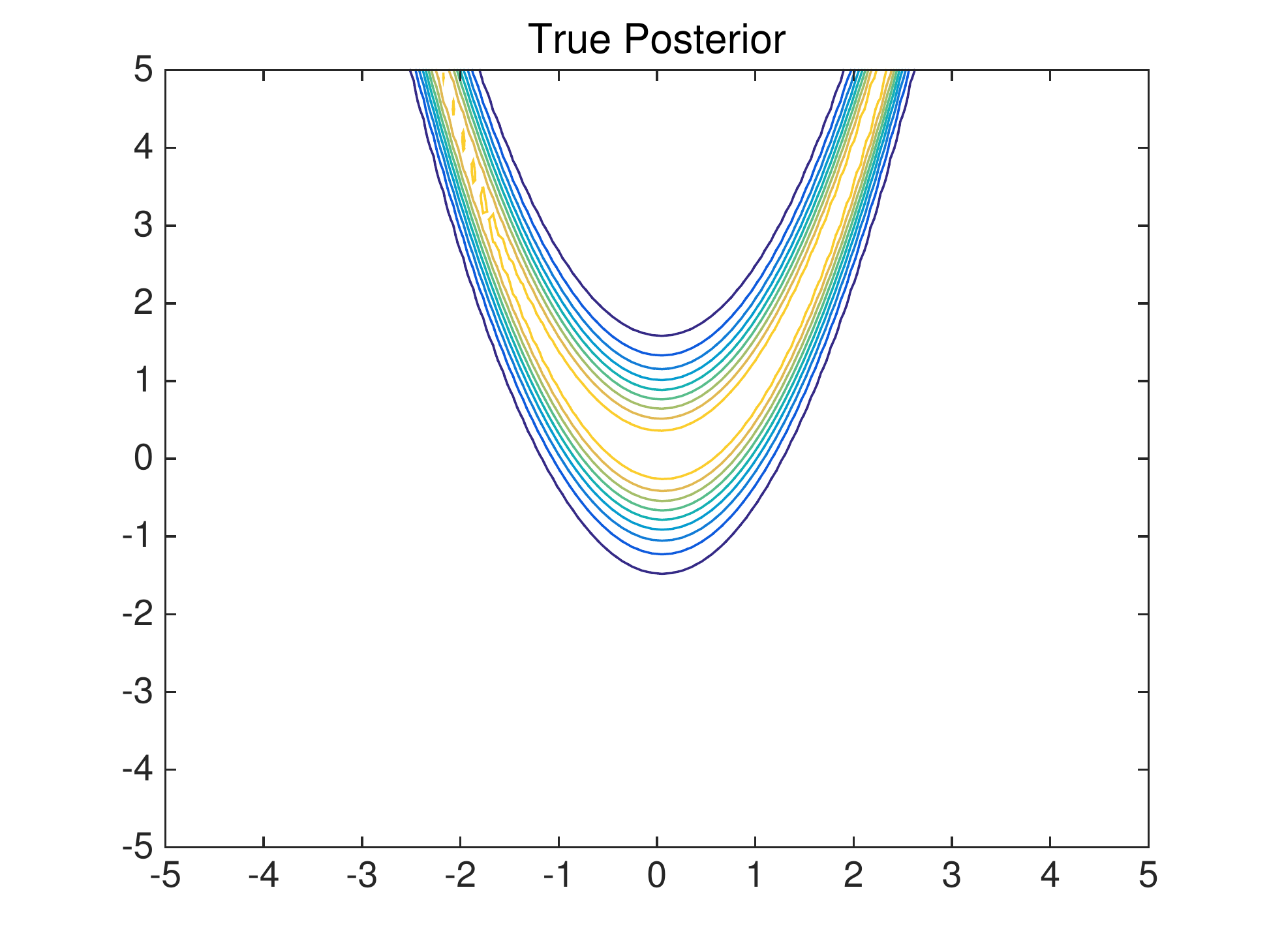} \includegraphics[width=.5\textwidth]{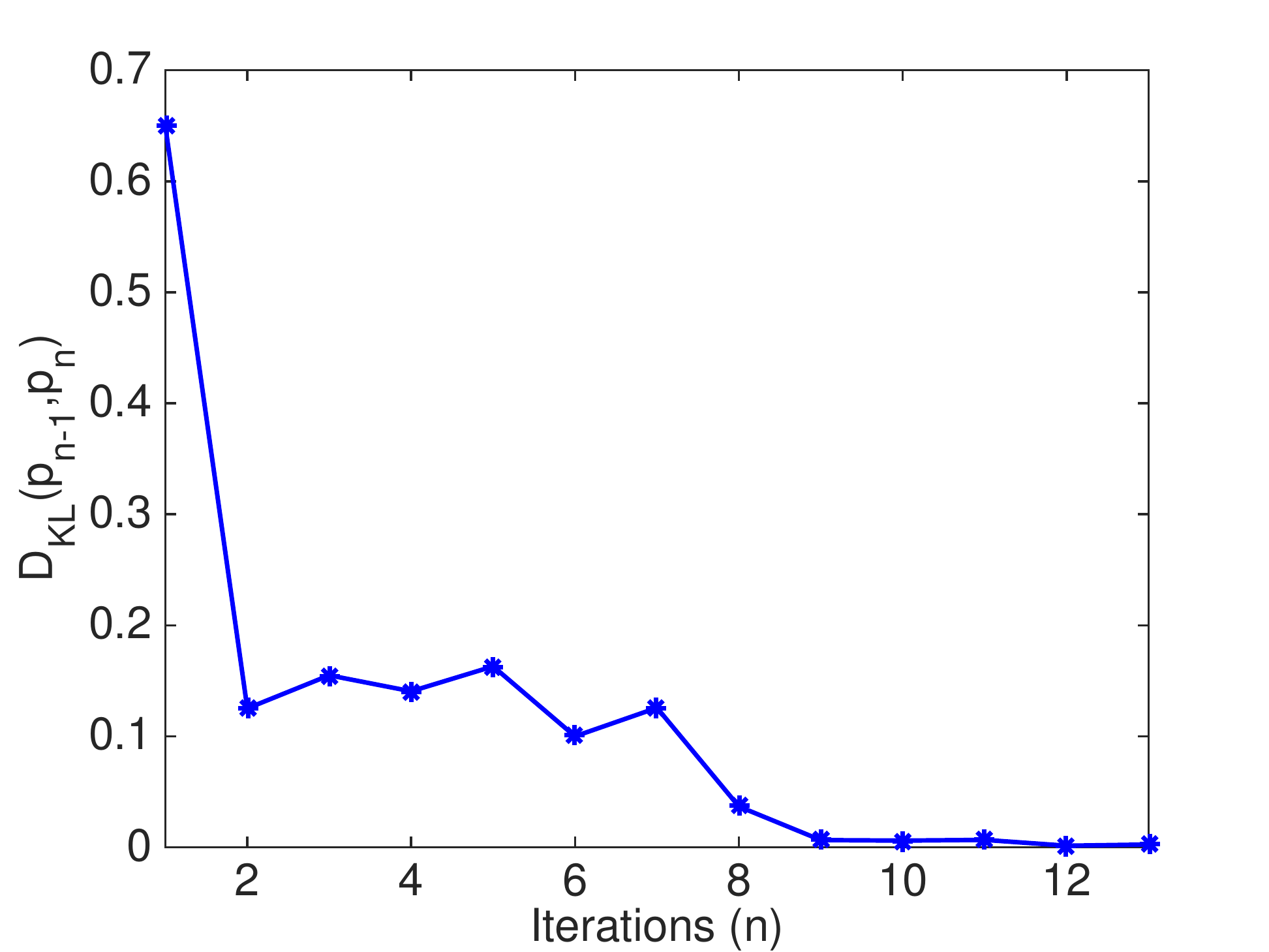}} 
\caption{Left: the true posterior distribution. Right: the KL distance between $p_{n-1}$ and $p_{n}$, plotted against the number of iterations.}\label{f:iteration}
\end{figure}

We first test our method on a two-dimensional mathematical example. The likelihood function is 
\begin{equation}
l(\-x)=\exp\left(-\frac1{100}(x_1-1)^2-(x_1^2-x_2)^2\right),
\end{equation}
which is the well-known Rosenbrock function,
and the prior $\pi(\-x)$ is a uniform distribution defined on $[-5,\,5]\times[-5,\,5]$. 
The resulting unnormalized posterior is shown in in Fig.~\ref{f:iteration} (left).
The function has a ``banana shape'',
and is often used as a test problem for Bayesian computation methods. 

\begin{figure}
\centerline{\includegraphics[width=.5\textwidth]{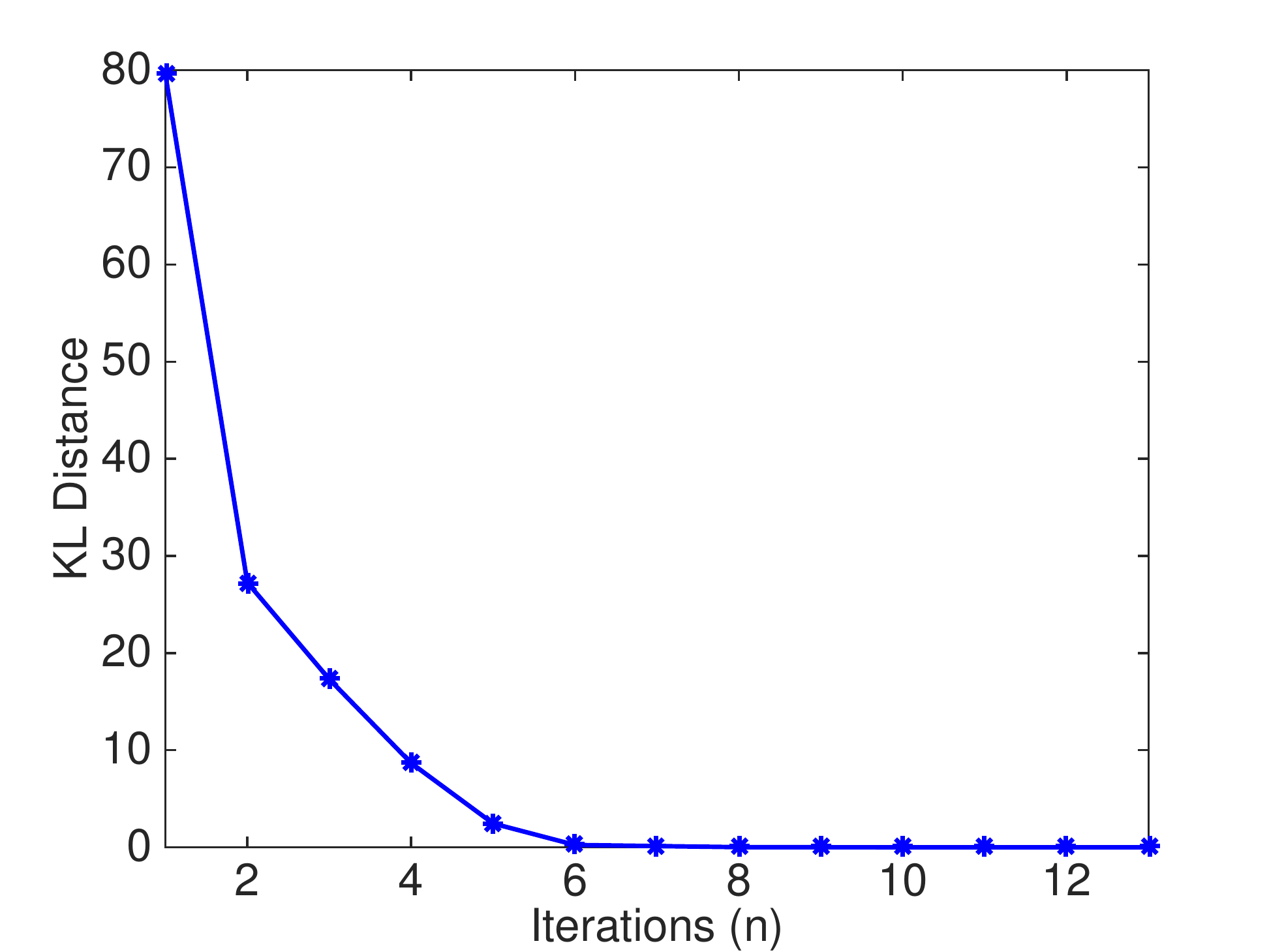} \includegraphics[width=.5\textwidth]{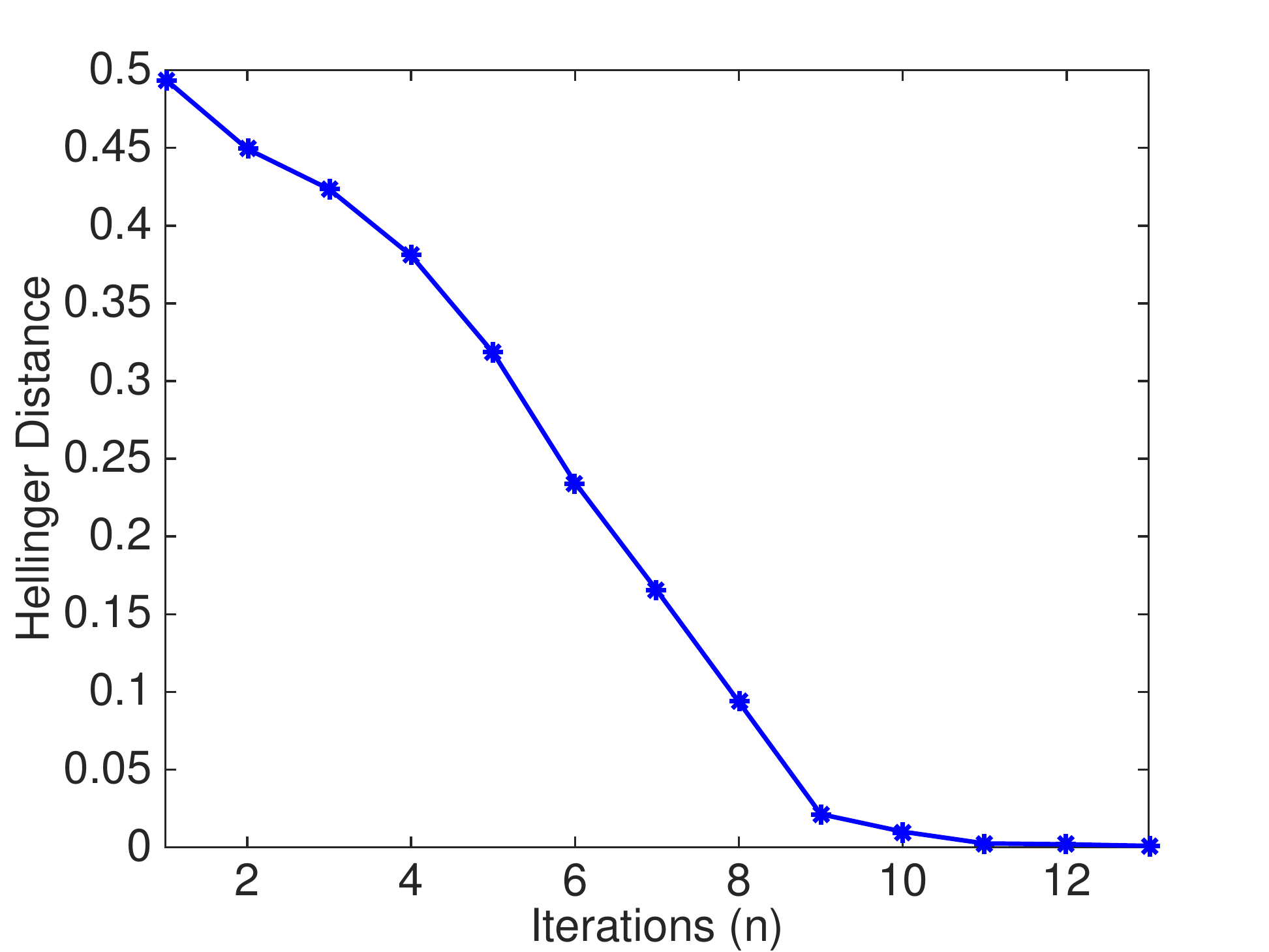}} 
\caption{ The KL (left) and the Hellinger (right) distances between the obtained approximation and 
the true posterior, plotted against the number of iterations.}\label{f:converge}
\end{figure}

\begin{figure}
\centerline{\includegraphics[width=.5\textwidth]{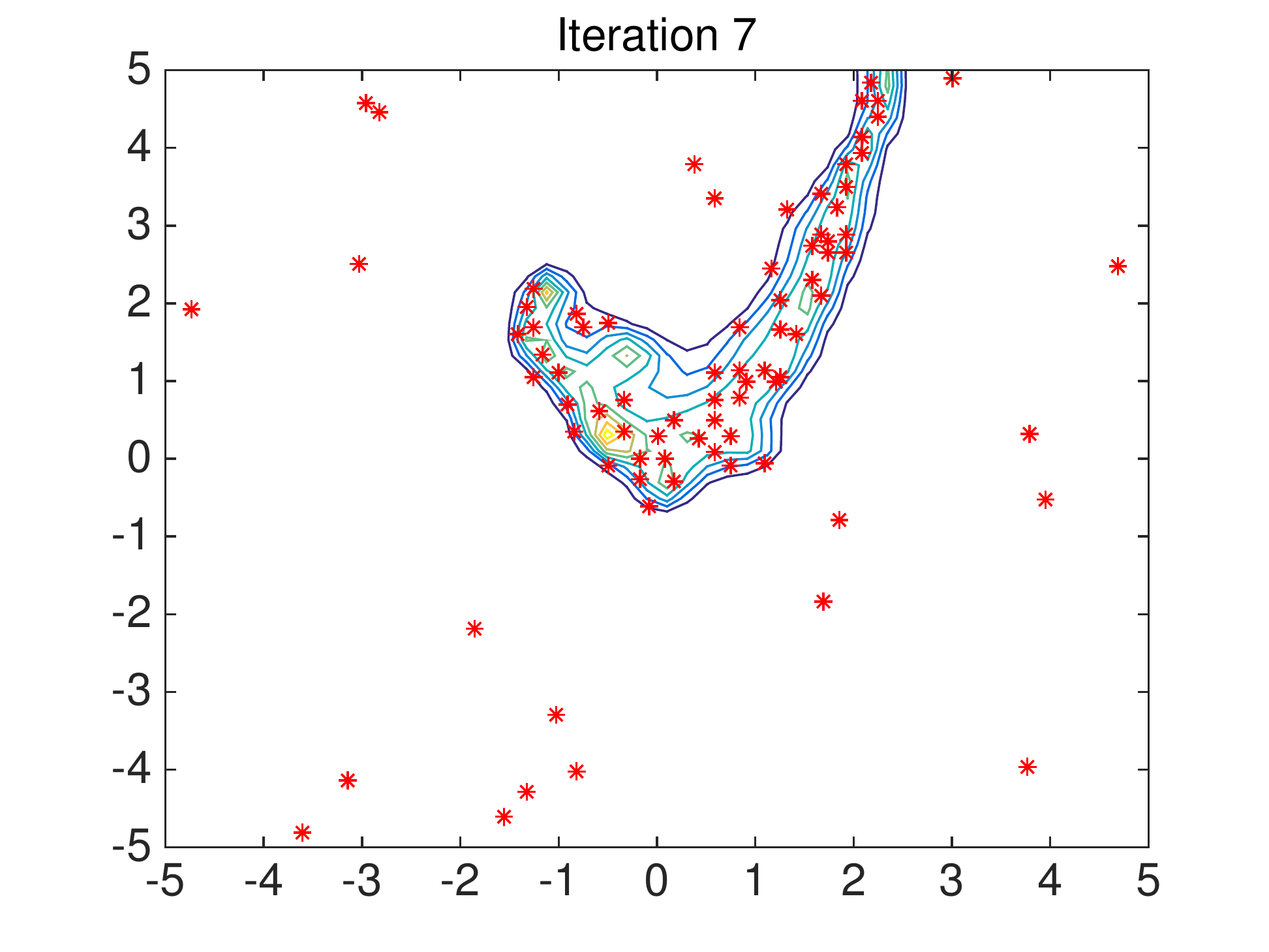}  
\includegraphics[width=.5\textwidth]{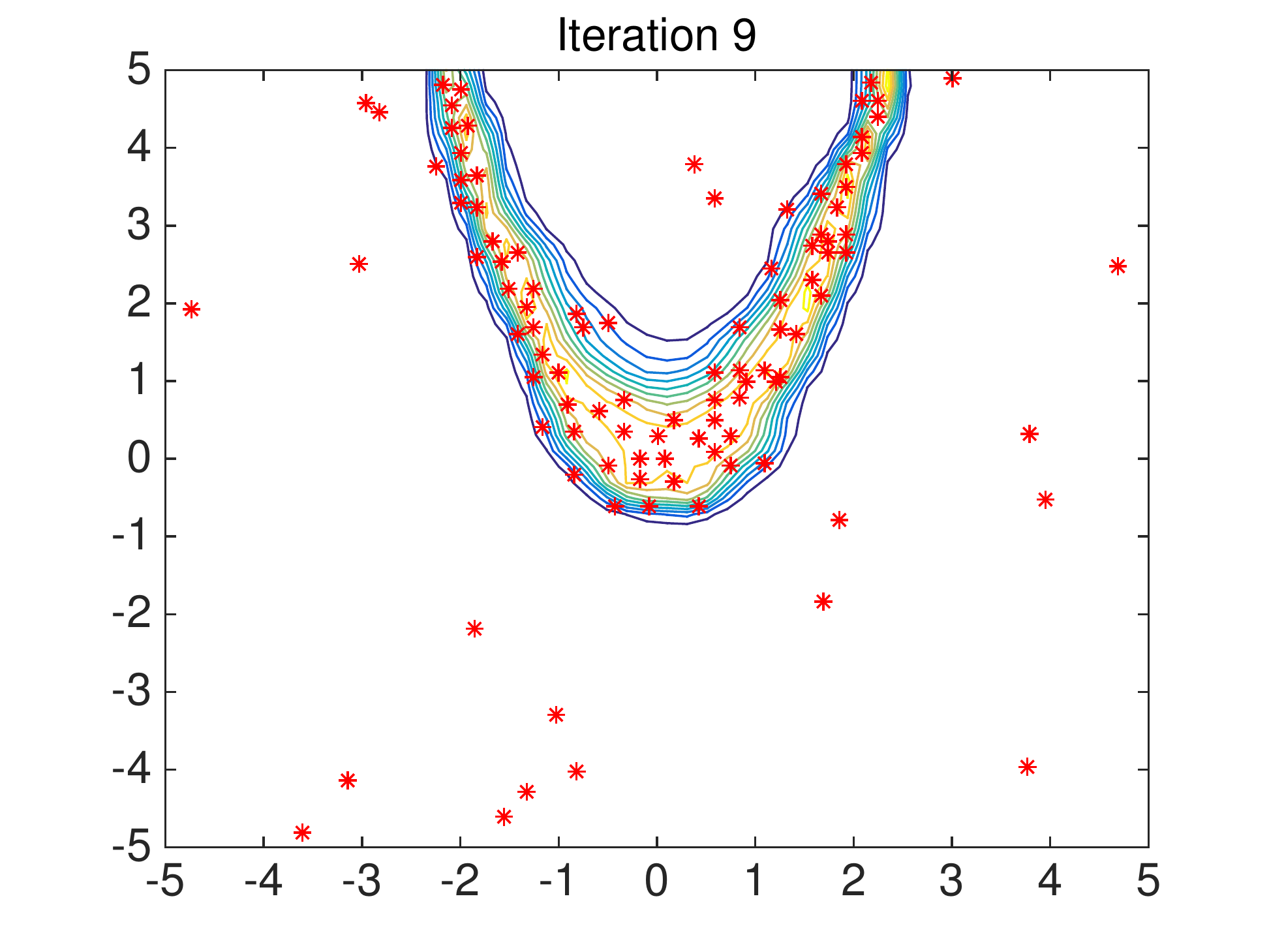}} 
\bigskip

\centerline{\includegraphics[width=.5\textwidth]{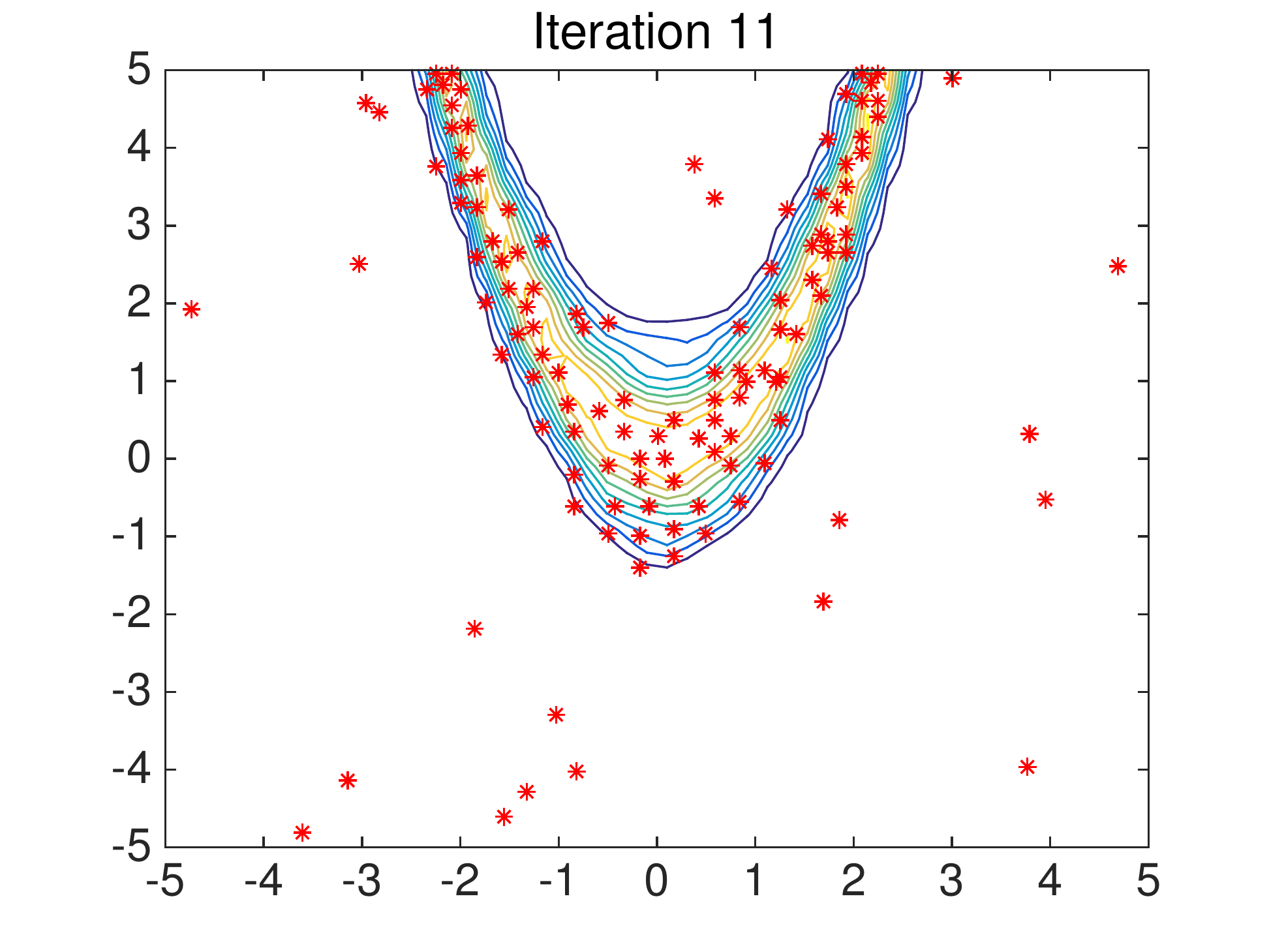}  
\includegraphics[width=.5\textwidth]{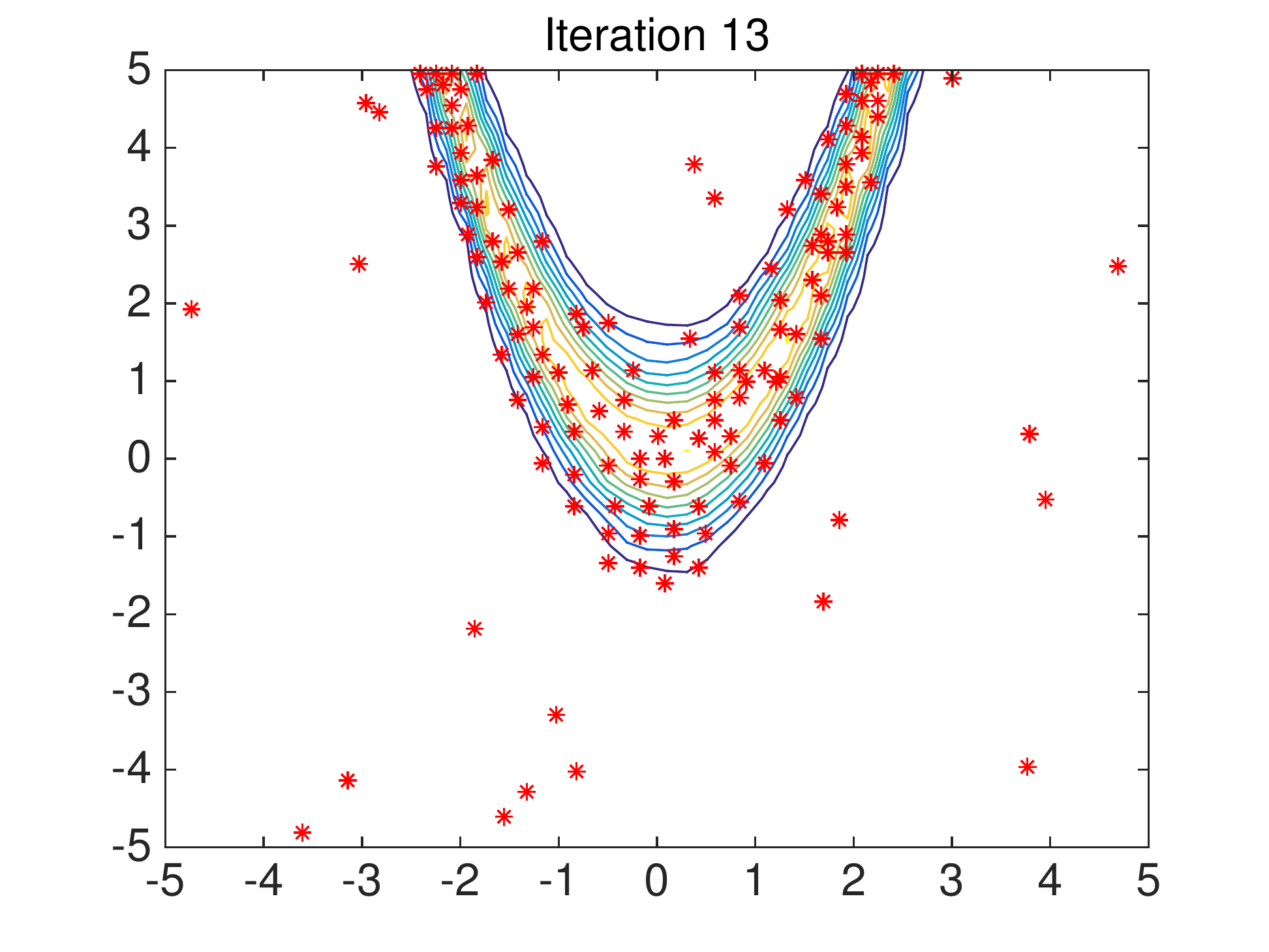}} 
\caption{The approximate posterior distribution obtained at the 7th, 9th, 11th and 13th iterations respectively. The red dots are the design points that have been used.}\label{f:pdfs}
\end{figure}

\begin{figure}
\centerline{\includegraphics[width=.5\textwidth]{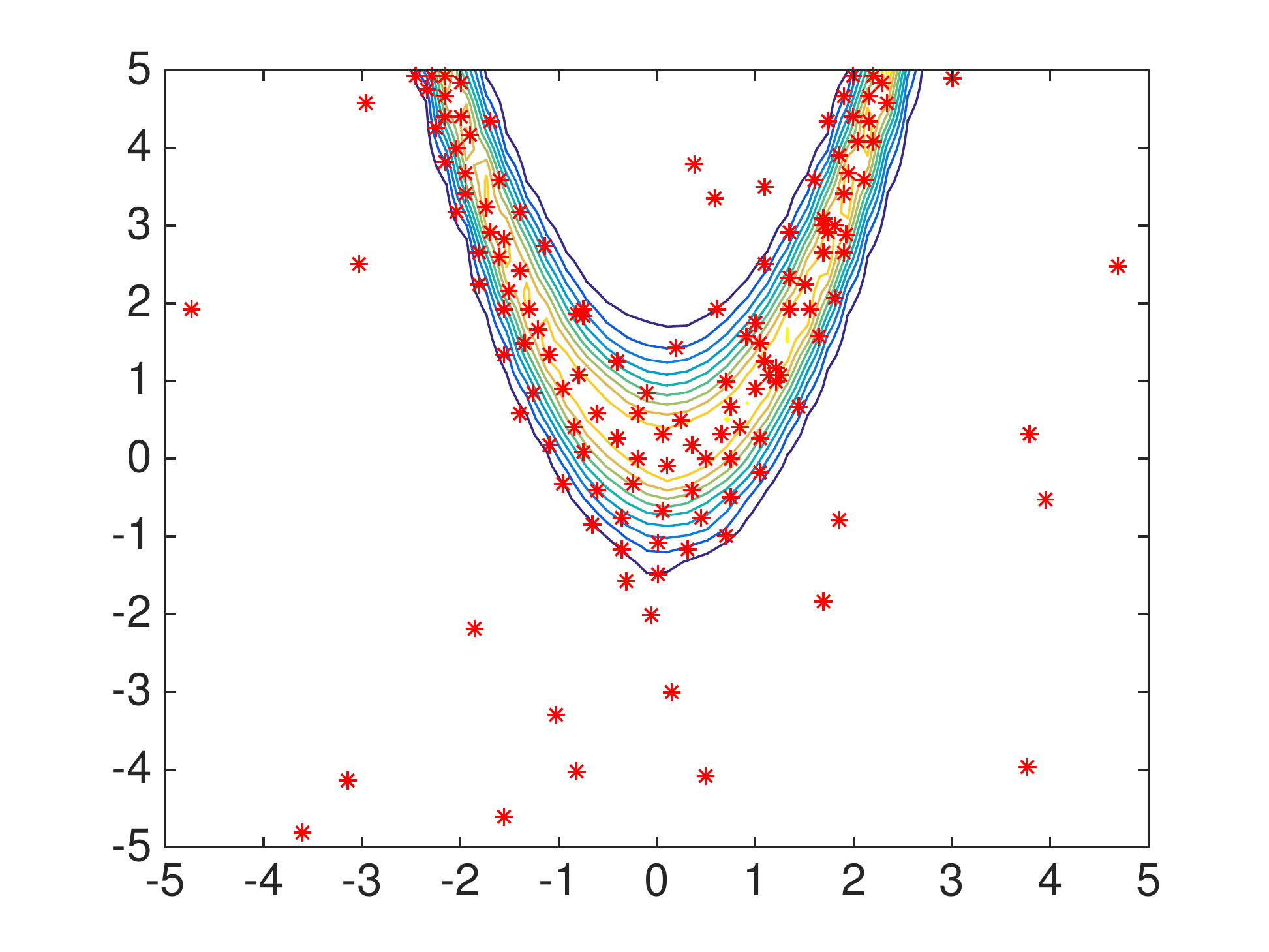} \includegraphics[width=.5\textwidth]{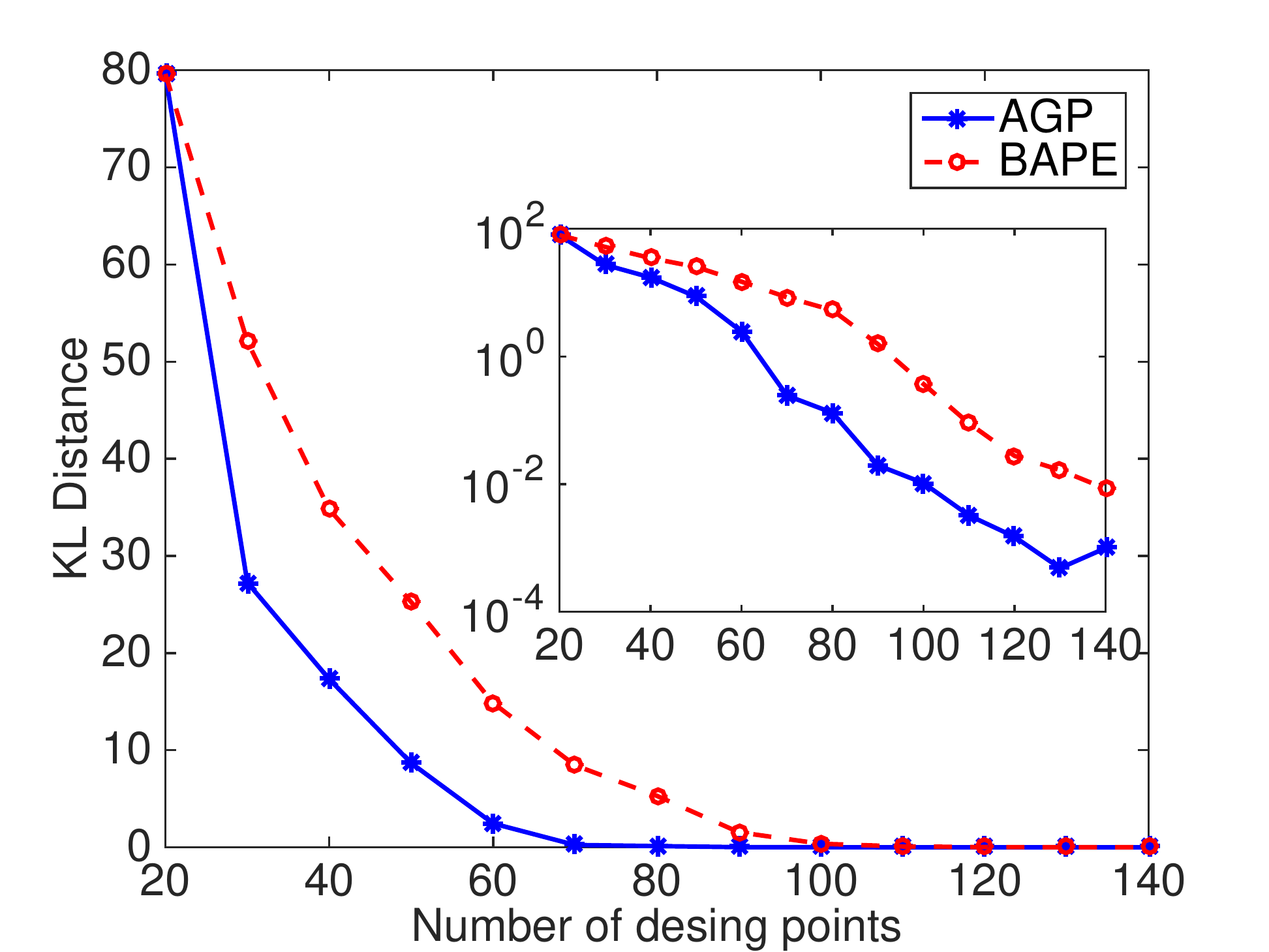}} 
\caption{Left: the GP approximation of the posterior distribution obtained with the BAPE method using 140 design points (red dots).
Right: the KL distance between the true posterior and the approximation computed with the AGP (solid line) and the BAPE (dashed line) methods, plotted again
the number of design points used; the inset is the same plot on a logarithmic scale.}\label{f:pdfs2}
\end{figure}

We now apply the proposed adaptive GP method to compute the posterior for this problem. 
In this example, we let $m_0 = 20$ and the samples in $S_0$ were randomly drawn according to the prior distribution. 
We also choose $m=10$: namely, $10$ new design points are computed in each iteration. 
In the algorithm, we need to sample from the approximate posterior distribution in each iteration, and here we draw $M=2\times10^4$ samples with the delayed rejection adaptive Metropolis algorithm (DRAM)~\cite{haario2006dram}. 
We reinstate that the $2\times10^4$ MCMC samples are generated from the approximate posterior distribution and thus it does not require to evaluate the
true likelihood function. 
We also set the parameters that specify the termination conditions to be $n_{\max}=100$,  $D_{\max}=0.01$ and $K=5$. 
The algorithm terminates with 13 iterations and totally 140 evaluations of the true likelihood function are used.
In Figs.~\ref{f:iteration} (right), we plot the KL difference in two consecutive iterations, 
which is used as one of our stoping criteria,  against the number of iterations. 
To illustrate the performance of our method, we use the KL distance and the Hellinger distance which is defined as, 
\[ D_H(p_1,p_2) = \frac12\int(\sqrt{p_1(\-x)}-\sqrt{p_2(\-x)})^2 d\-x,\]
to quantify the difference between the computed approximation and 
the true posterior.
We plot  the KL (left) and the Hellinger (right) distances between the approximate posterior and the true posterior distribution in   Figs.~\ref{f:converge}. 
It can be seen from the figures that, the computed approximation converges very well to the true posterior in terms of  both 
distance mesures, as the iteration proceeds. 
We then plot the approximate posterior obtained in the 7th, 9th, 11th and 13th iterations in Figs.~\ref{f:pdfs}, in which we can visualize how the quality of the approximation increases as  the iterations proceed. 
In each of the plots, we also show the design points (red dots) that have been used up to the given iteration. 
As a comparison, we also compute the GP approximation of the posterior with 
the aforementioned Bayesian active posterior estimation (BAPE) method developed in \cite{kandasamy2015bayesian}. 
In particular, we implement the BAPE method using totally $140$ design points
and this way it matches the number of design points of our method. 
The results are shown in Figs~\ref{f:pdfs2}.
The figure on the left shows  the posterior distribution computed with all the140 design points (corresponding to the 13th iteration in our method),
and as one can see, the BAPE method can also obtain a good approximation of the posterior distribution.
To compare the performance of the two methods, we compute 
the KL divergence between the true posterior and the approximation obtained with different numbers of design points
by the BAPE and our adaptive GP (AGP) methods.
We plot the KL distance against the number of design points in Fig.~\ref{f:pdfs2} (left).
One can see from the figure that,  with the same number of design points, 
the approximate posterior obtained by the proposed AGP method is significantly closer to the true posterior than the results of the BAPE method.  

\subsection{Genetic toggle switch}
We now apply the proposed method to a real-world inference problem. Namely, we consider the kinetics of a genetic toggle switch,  which was first studied in \cite{gardner2000construction} and later numerically 
investigated in \cite{marzouk2009stochastic}. The toggle switch consists of two repressible 
promotors arranged in a mutually inhibitory network: promoter 1 an promoter 2.
Either promoter transcribes a repressor
for the other one, and moreover, either repressor may be induced by an external chemical or thermal signal.
Genetic circuits of this form can be modeled by the following differential-algebraic
equation system~\cite{gardner2000construction}:
\begin{subequations}
\label{e:gt}
\begin{eqnarray}
\frac{du}{dt}&=& \frac{\alpha_1}{1+v^\beta}-u,\\
\frac{dv}{dt}&=&\frac{\alpha_2}{1+w^\gamma}-v,\\
w&=&\frac{u}{1+([\mathrm{IPTG}]/K)^\eta}.
\end{eqnarray}
\end{subequations}
In the equations above, $u$ and $v$ are respectively the concentration of repressors 1 and 2;
$\alpha_1$ and $\alpha_2$ are the effective rates of synthesis of the repressors; $\gamma$ and $\beta$
represent cooperativity of repression of the two promotors; 
and [IPTG] is the concentration of IPTG, the chemical compound that induces the switch. 
Parameters $K$ and $\eta$ describe binding of IPTG with the first repressor. 
For more details of the model, we refer to \cite{gardner2000construction}. 
\begin{table}[!htb]
\center
\ {
\medskip

\begin{tabular}{lcccccc}
\hline %
&$\alpha_1$& $\alpha_2$&  $\gamma$ & $\beta$ & $\eta$ & $K$\\
\hline %
$n$ & $[120, 200]$ & $[15.0, 16.0]$ & $[2.1, 2.9]$& $[0.85, 1.15]$&$[1.3, 2.7]$ & $[2.3, 3.7]\times10^{-5}$ \\
  \hline
 \end{tabular}
}
\caption{The prior domains of the parameters.}
\label{tb:domains}
\end{table}

The experiments are performed with several selected values of [IPTG]: $1\times10^{-6}, 5\times10^{-4}, 7\times10^{-4}, 
1\times10^{-3}, 3\times10^{-3}, 5\times10^{-3}$ respectively,
and for each experiment, the measurement of $v$ is taken at $t=10$. 
The goal is to infer the six parameters \[\-x=[\alpha_1, \alpha_2, \gamma, \beta,  \eta, K],\]
from the measurements of $v$. 
We use synthetic data in this problem, and specifically we assume that the true values of the parameters are
\[\-x_\mathrm{true}=[143, 15.95, 2.70, 0.96, 2.34, 2.70\times10^{-5}].\]
The data is simulated using the model described by Eqs.~\eqref{e:gt} with
the true parameter values and measurement noise is then added to the simulate data. 
The measurement noise here is assumed Gaussian and zero-mean, with a variance $\sigma^2$.
In the numerical experiments, we consider a large noise case where $\sigma^2=5\times10^{-4}$ and a small noise case where $\sigma^2 = 1.25\times10^{-4}$.  
We assume that the priors of the six parameters are all uniform and independent of each other, where the domains of the uniform priors are given in Table~\ref{tb:domains}. 

{   We want to use this example to make a detailed comparison of the proposed AGP algorithm with some other popular methods.
Thus, we employ four different methods to compute the posterior distribution in this example: the direct MCMC algorithm with the true likelihood function, the proposed AGP algorithm, 
 the BAPE method~\cite {kandasamy2015bayesian},
  and the spectral likelihood expansion~(SLE) method~\cite{nagel2016spectral} which constructs 
the gPC surrogate for the likelihood function using non-intrusive approaches.}

{We first consider the large noise case. 
We  draw $3\times10^{5}$ samples from the true posterior distribution with a DRAM algorithm and use the results as the reference posterior distribution. 
We then apply the AGP method to approximate the posterior distribution, where we use $m_0=50$ initial design points randomly drawn from the prior and
$m=50$ design points in each iteration. 
We also choose the termination parameters to be $n_{\max}=100$, $D_{\max}=0.05$ and $K=5$. 
The algorithm terminates in 18 iterations, resulting in totally 
950 evaluations of the true likelihood function. }
We note that each evaluation of the likelihood function involves a full simulation of the underlying model described by Eqs.~\eqref{e:gt}. 
After obtaining the approximate posterior distribution, we draw $3\times10^{5}$ samples from it using a DRAM MCMC simulation.
We then compute the approximate posterior with the BAPE method using also 950 likelihood evaluations, and draw $3\times10^{5}$ samples from it using a DRAM MCMC simulation.
Finally we approximate the posterior with the SLE-gPC method where the gPC expansion coefficients are computed using the least square method with design points determined by
the Sobol sequences based quasi Monte Carlo (QMC) method. The gPC degree is automatically by the algorithm using the leave-one-out (LOO) cross validation,
and in the QMC scheme, we set the number of design points to be 950. 

\begin{figure}
\centerline{\includegraphics[width=1\textwidth]{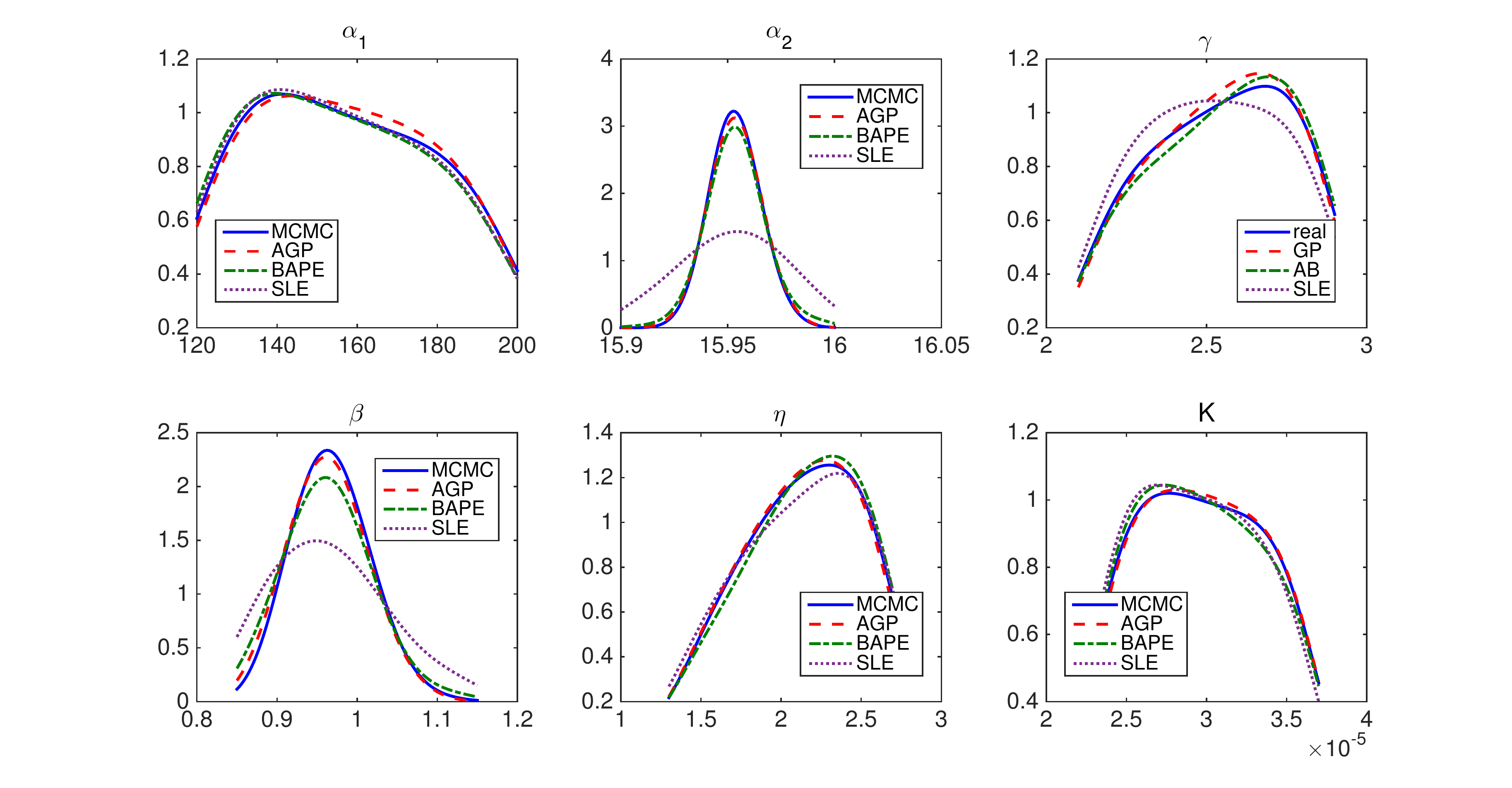}} 
\caption{The genetic toggle example: large noise case. The marginal distributions of the six parameters, computed with the four different methods.}\label{f:ex2ln}
\end{figure}

We now compare the results of these methods. First we estimate the posterior distributions of the six parameters by all the four methods
and show the results in Fig.~\ref{f:ex2ln}. 
One can see from the figure that, the distributions computed by both the BAPE and the AGP methods are rather close to those of direct MCMC (which are regarded as the true posteriors),
while the results of SLE-gPC deviate evidently from the MCMC results, especially for the two parameters $\alpha_2$ and $\beta$.
As for the comparison of BAPE and AGP,  the figures show that both methods can produce reasonably good approximations of the posterior distributions in this case. So 
for a quantitive evaluation fo the performance of the  methods, we compute the KLD from the approximate posterior distributions to the true posterior densities, and show the results 
in Table~\ref{tb:kldln}. We can see from the table that, the posterior distributions computed by the 
AGP method are closer (in terms of KLD) to the true posteriors than the other two methods for all the 6 parameters. 
%

\begin{figure}
\centerline{\includegraphics[width=1\textwidth]{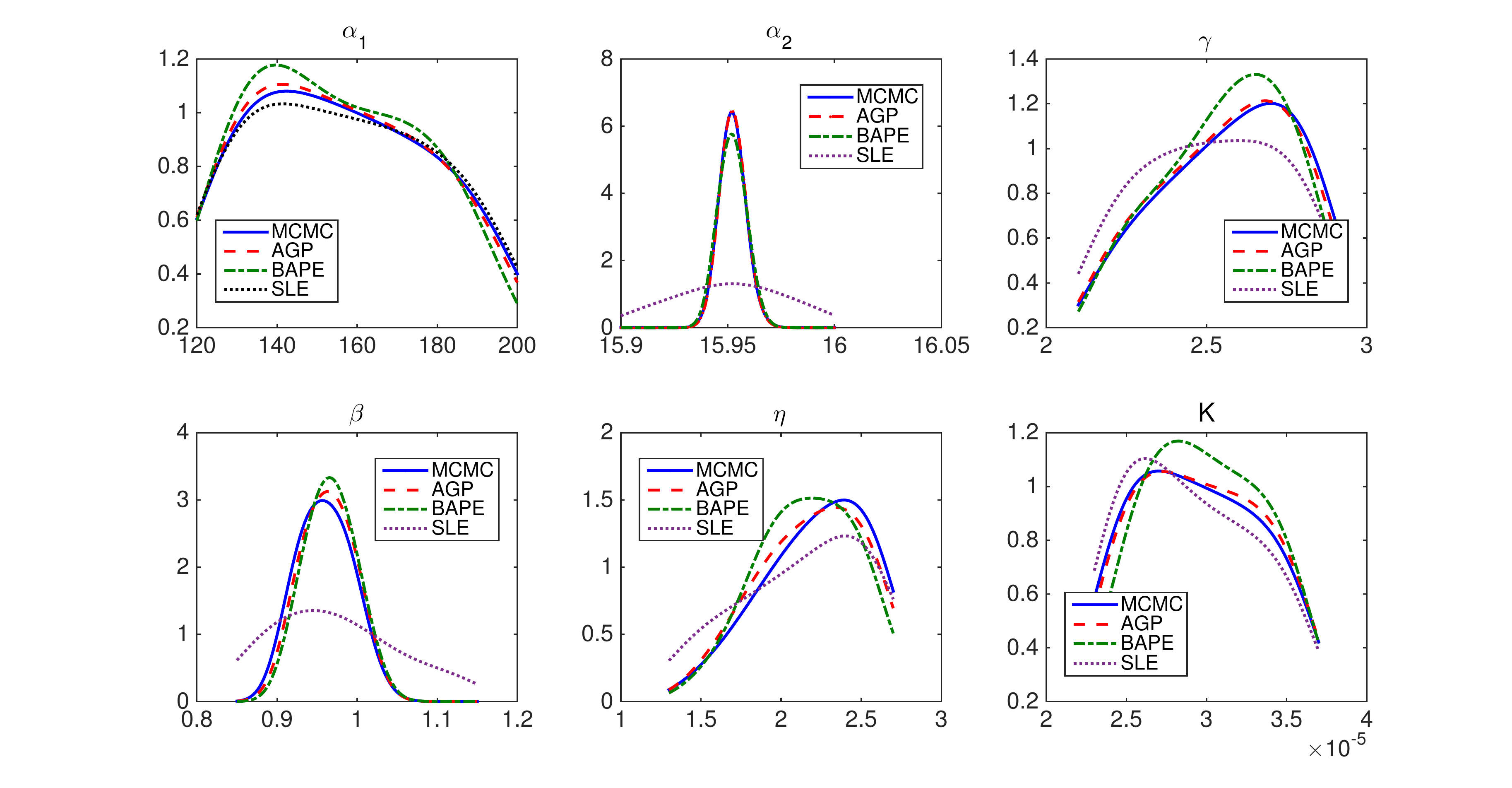}} 
\caption{The genetic toggle example: small noise case. The marginal distributions of the six parameters, computed with the four different methods.}\label{f:ex2sn}
\end{figure}

We then consider the small noise case where we use the same implementation configurations as the large noise case. 
In this case, the AGP algorithm uses 1200 true likelihood evaluations, 
and as before, we also compute the posterior using the SLE and the BAPE methods with the same number of true likelihood evaluations. 
We then compute the posterior directly with $3\times10^5$ MCMC samples and use the results as the true posterior. 
In Figs.~\ref{f:ex2sn} we compare the marginal posterior distributions compared with the four methods. 
Similar to the large noise case,  the figures show that the results of the SLE-gPC are of very low accuracy, while both the AGP and the BAPE methods yield rather good results. 
Once again, we show in Table~\ref{tb:kldln} the KLD from the marginal posterior distributions computed with the three approximate methods to 
the true posterior (those computed by the direct MCMC). These quantitive comparison results indicate that, the AGP method yields better results than 
the other two methods in terms of the KLD. Thus, we can conclude that our AGP method has the best performance in both the large and the small noise cases.

%

\begin{table}[!htb]
\center
\medskip
\begin{tabular}{l|ccccccc}
\hline
  &method &$\alpha_1$& $\alpha_2$&  $\gamma$ & $\beta$ & $\eta$ & $K$\\
\hline
&AGP & $1.5\times10^{-4}$ &    $0.0069 $ &  $0.0020$ &  $ 0.014$  & $ 0.0041$ & $0.0075$ \\
\cline{2 -8} %
KLD &BAPE & $4.5\times10^{-3}$ & $0.043$ & $0.0022$& $0.039$&$0.0095$ & $0.013$ \\
  \cline{2 -8} %
 &SLE & $1.8\times10^{-3}$ & $0.44$ & $0.0012$& $0.18$&$0.0078$ & $0.0013$ \\
 \hline
 \end{tabular}
\caption{The large noise case: the KLD between the marginal posterior distributions computed with the three approximate methods 
and those computed with standard MCMC.}
\label{tb:kldln}
\end{table}

%

\begin{table}[!htb]
\center
\medskip
\begin{tabular}{l|ccccccc}
\hline
  &method &$\alpha_1$& $\alpha_2$&  $\gamma$ & $\beta$ & $\eta$ & $K$\\
\hline
&AGP & $0.0032$ &    $0.015 $ &  $0.0039$ &  $ 0.012$  & $ 0.0096$ & $0.014$ \\
\cline{2 -8} %
KLD &BAPE & $0.015$ & $0.036$ & $0.028$& $0.057$&$0.0075$ & $0.074$ \\
  \cline{2 -8} %
 &SLE & $0.0035$ & $1.1$ & $0.021$& $0.46$&$0.050$ & $0.010$ \\
 \hline
 \end{tabular}
\caption{The small noise case: the KLD between the marginal posterior distributions computed with the three approximate methods 
and those computed with standard MCMC.}
\label{tb:kld2}
\end{table}

\subsection{  The human body sway problem}
Finally, we apply the proposed method to a human body sway problem. 
This problem has received considerable attention as the body sway may provide information about the physiological status of a person~\cite{Tiet2017Bayesian}. 
Several mathematical models have been proposed to describe the sway motion,  and here we consider the single-link inverted pendulum (SLIP) model proposed  in \cite{Asai2009A},
which assumes that the body is maintained in an upright position by an active and a passive proportional-derivative controller. 
  
%
%

Specifically, the SLIP model is given by the following  stochastic delay differential equation (SDDE)~ \cite{Asai2009A}:  
\begin{equation}
I\ddot{\theta}(t)=mgh\theta(t) - [K\theta(t) + B\dot{\theta}(t) + f_P(\theta(t-\Delta)) + f_D(\dot{\theta}(t-\Delta))] +  \xi(t). 
\label{eq:3-2}
\end{equation}
In this equation, $I$ is the moment of inertia of the body, $\theta$ is  the tilt angle ($\dot{\theta}$ and $\ddot{\theta}$ are its first and second derivatives respectively),
 $m$ denotes the body mass, $g$ is the gradational acceleration, 
 $h$ is the distance between 3D center-of-mass (COM) and the ankle joint, and $\xi$ is a zero-mean Gaussian noise with variance $\sigma^2$. 
 $K$ and $P$ are the passive stiffness and passive damping parameters, and
 $f_P(\theta(t-\Delta))$ and $f_D(\dot{\theta}(t-\Delta))$ are active stiffness and active damping terms where $\Delta$ is the time delay. 
 We now specify the active stiffness $f_P({\theta}(t-\Delta))$ and the active damping $f_D(\dot{\theta}(t-\Delta))$.
 We first define two functions $c_1(\theta(t-\Delta)) = \theta(\dot{\theta}(t-\Delta) - a_s\theta(t-\Delta))$
 and $c_2(\theta(t-\Delta))=\theta(t-\Delta)^2+(\dot{\theta}(t-\Delta))^2$. 
 We then have, 
\begin{equation}
f_P(\theta(t-\Delta)) =\left \{ \begin{array}{ll}
 P\theta(t-\Delta),\quad \mbox{if} \,\,c_1(\theta(t-\Delta))>0 \,\,\mbox{and} \,\,c_2(\theta(t-\Delta))>r^2;\\
0,\quad \mbox{otherwise;}
\end{array}
\right.
\end{equation}
and
\begin{equation}
f_D(\dot{\theta}(t-\Delta)) =\left \{ \begin{array}{ll}
 D\dot{\theta}(t-\Delta)\quad \mbox{if} \,\,c_1(\theta(t))>0 \,\,\mbox{and} \,\,c_2(\theta(t))>r^2;\\
 0,\quad \mbox{otherwise;}
\end{array}
\right.
\end{equation}
where $r$ is the radius of the ``quiet zone'' (active control is off).
The slope $a_s$ depends on the level of control, $C_{ON}$, as
 $a_s=-\tan(\pi(C_{ON}-0.5))$.
In this model, five key parameters ($P$, $D$, $\Delta$, $\sigma$, and $C_{ON}$) can not be measured directly and need to be inferred from the body sway measurements, while the other parameters can either be measured or specified in advance~\cite{Tiet2017Bayesian}.  
The COM signal,
\begin{equation}
\mbox{COM}(t) = h\sin(\theta(t)),
\end{equation}
is measured and used to infer the five unknown model parameters. 
The absolute value of the COM amplitude, velocity, acceleration and the power spectral density (PSD) are extracted from the signal. The mean, variance, skewness and kurtosis of each physical quantity are calculated as the data $\-y$ (a $16$-dimensional vector) to infer the model parameters:
$\-x =(P, \,D, \,\Delta, \,\sigma,\, C_{ON})$.
Computing the posterior in this problem is rather challenging as the likelihood function $p(\-y|\-x)$ is not available, 
which make the standard Bayesian inference computation methods such as the MCMC algorithms infeasible. 
 In  \cite{Tiet2017Bayesian}, the parameters were inferred with the approximate Bayesian computation (ABC) method which does not use the likelihood function. 
 Here we compute the posterior with the proposed method. 
 In particular, we compute the likelihood function using the following procedure. For a given parameter value $\-x$, we perform a Monte Carlo simulation for the SDDE model~\eqref{eq:3-2} with a given number of samples.  
 For each $y_i$ for $i=1...16$, we estimate resulting conditional density function $p_i(y_i|\-x)$ with the kernel density estimation method, and then we take the likelihood function to be 
 \[ p(\-y|\-x) = \prod_{i=1}^{16} p_i(y_i|\-x).\]
We note that a single evaluation of the likelihood function requires to repeatedly simulate Eq.~\eqref{eq:3-2} a large number of times, which renders the evaluation highly intensive.   

\begin{figure}
\centerline{\includegraphics[width=1\textwidth]{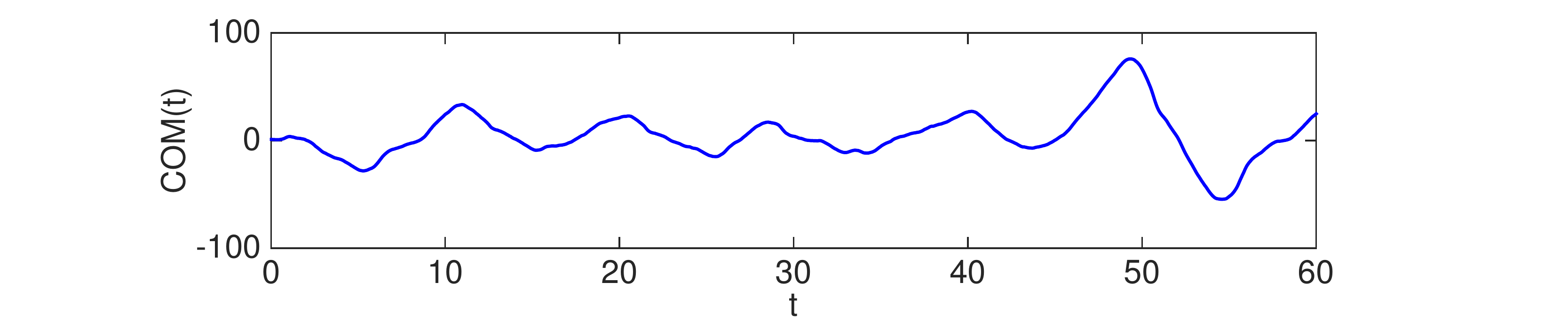}} 
\caption{The simulated COM signal.}\label{f:COM}
\end{figure}
\begin{figure}
\centerline{\includegraphics[width=.9\textwidth]{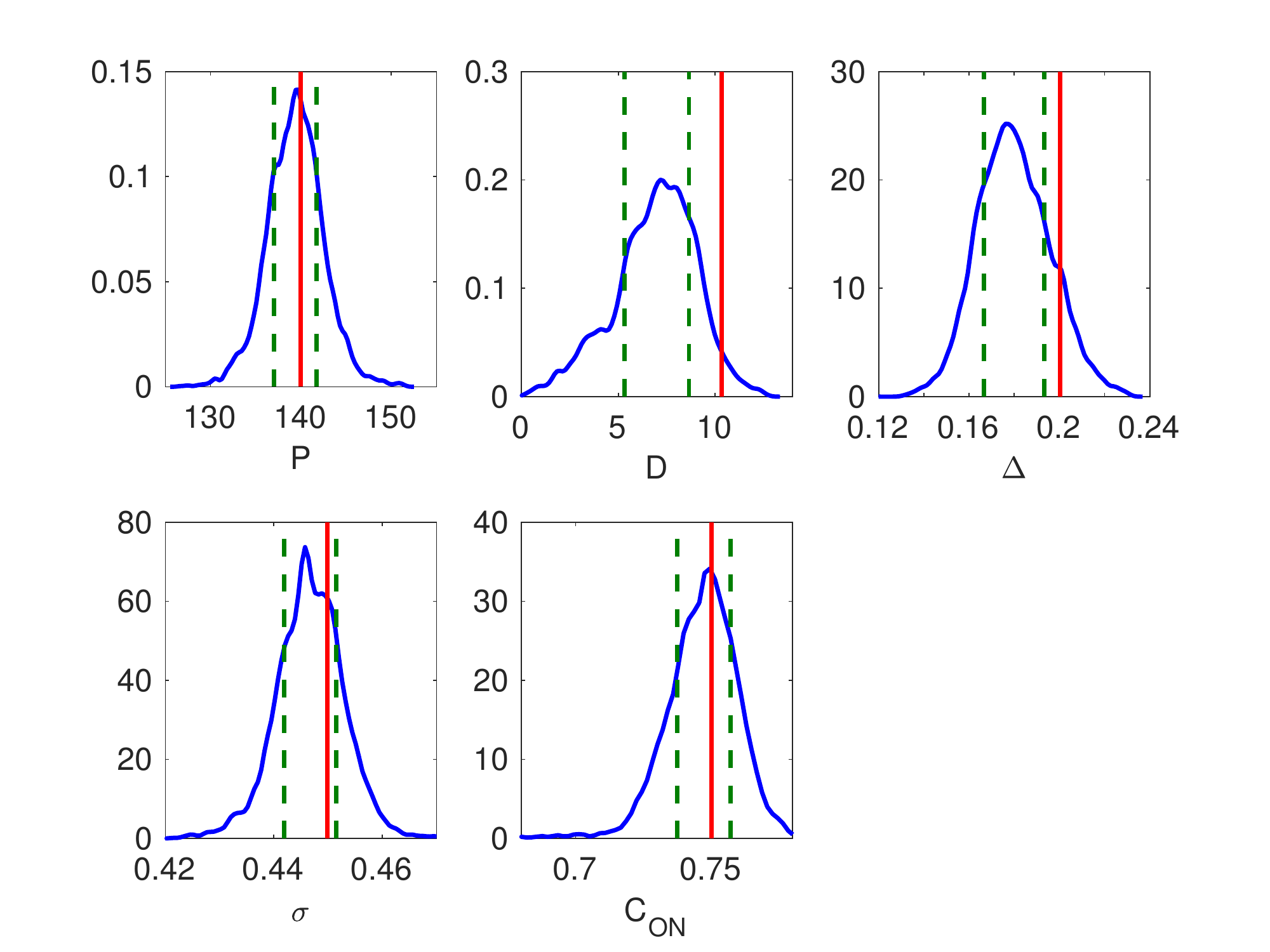}} 
\kern-2\bigskipamount
\caption{\small The posteriors of the parameters in the SLIP model, computed by the BAPE method.} 
Also shown in the figures are the 60\% confidence interval (dashed vertical lines) and the true parameter values (solid vertical lines). 
\label{f:ex3post_bape}
\medskip

\centerline{\includegraphics[width=.9\textwidth]{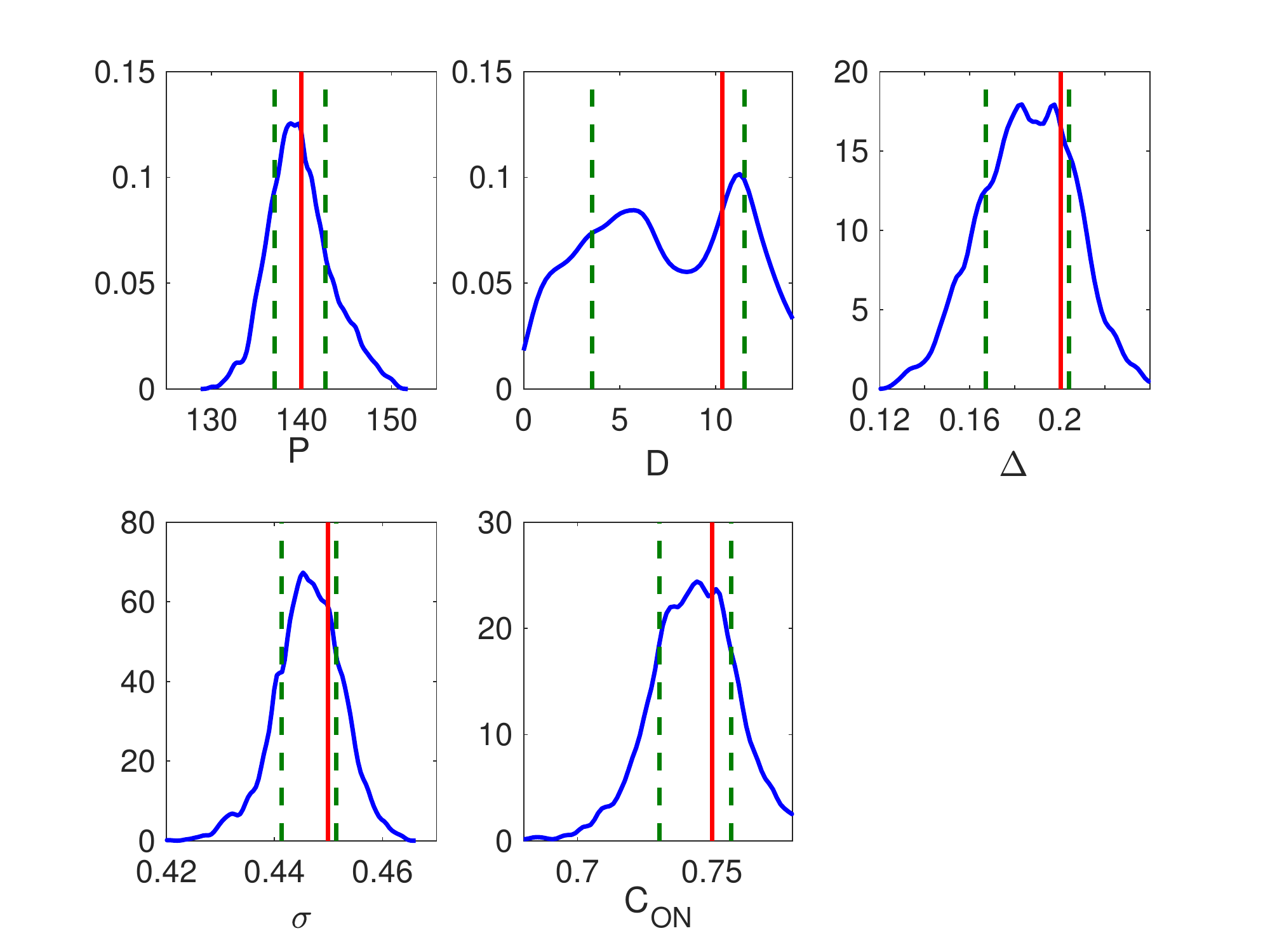}} 
\kern-2\bigskipamount
\caption{\small The posteriors of the parameters in the SLIP model, computed by our AGP method. Also shown in the figures are the 60\% confidence interval (dashed vertical lines) and the true parameter values (solid vertical lines). }\label{f:ex3post_agp}
\end{figure}

In the numerical experiments, we use simulated data, and in particular the true parameters values are set to be $P= 145 Nm/rad$, $D=10 Nms/rad$, $\Delta=0.2s$, $\sigma=0.45Nm$ and $C_{ON}=0.75$,
and the other parameter values are $g=9.81m/s^2$, $m = 68kg$, $h = 0.87m$, $I = mh^2$,  $K = mgh\times 0.8Nm/rad$,  $B = 4 Nms/rad$ and $r = 0.004rad\hbox{-}rad/s$. 
The COM signal generated from the model Eq.~\eqref{eq:3-2} with these parameter values is shown in Fig.~\ref{f:COM}.
In the inference, we impose a uniform distribution on each of the five parameters on the following intervals: $P\in[80,\,160]$, $D\in[0.05,\,30]$, $\Delta\in[0.05,\,0.5]$, $\sigma\in[0.05,\,0.6]$, and $C_{ON}\in[0.05,\,0.85]$. 
Moreover,  for each evaluation of the likelihood function  $\pi(\-y|\-x)$, we use 10,000 simulations of Eq.~\eqref{eq:3-2} , 
and as a result a direct MCMC simulation of the posterior distribution is computationally infeasible. 
We apply our AGP method to compute the posterior distribution, and the algorithm parameters are the same as those in the second example.
The algorithm terminates in 14 iterations and so total number of true likelihood evaluations is 750. 
As a comparison, we also perform the BAPE method with the same number of true likelihood function evaluations. 
To compare the performance of the two methods,  we plot the posterior marginals of the model parameters computed by BAPE in Fig.~\ref{f:ex3post_bape}, 
and those computed by AGP in Fig.~\ref{f:ex3post_agp}. 
We also show the true parameter values as well as the 60\% confidence interval in the figures.
Here one can see that, for all the posteriors computed with the AGP method, the true parameter values fall in the 60\% confidence intervals,
while for the results of the BAPE method, the true values of $D$ and $\Delta$ fall outside of the 60\% confidence intervals,
which suggest that the posteriors computed by the AGP method may be more accurate and reliable that those by BAPE. 


\section{Conclusions}\label{sec:conclusions}


In summary, we have proposed an algorithm to construct GP based approximation for the un-normalized posterior distribution. 
The method expresses the un-normalize posterior as a product of an approximate posterior density and an exponentiated GP model,
and an adaptive scheme is presented to construct such an approximation. We also provide an active learning method that uses 
maximum entropy as the selection criterion to determine the sampling points. With numerical examples, we show that the method can obtain
a rather good approximation of the posterior with a limited number of evaluations of the likelihood functions. 
We believe the proposed method can be useful in a wide range of practical Bayesian inference problems
where the likelihood function are difficult or expensive to evaluate. 

Several issues of the proposed algorithm deserve further studies. 
First, while our numerical experiments  illustrate that the algorithm may converge in these examples,  a rigorous convergence analysis  of the algorithm is still lacking.  
Secondly, for a posterior distribution with unbounded domain, the resulting approximation may become improper, and thus certain modifications
of the algorithm may be needed to address the issue.
{  
Finally we note that selecting a good Kernel function for the GP model is a very important issue for GP-based methods, and 
to this end, a very interesting question is  how to choose kernel functions that are specifically suitable for the log-posteriors.}
We plan to study these issues in future works.

\subsection*{Acknowledgments}
The work was partially supported by the National Natural Science Foundation of China under grant number 111771289. 

\bibliographystyle{plain}
\bibliography{ak.bib}

\end{document}